\documentclass[10pt,conference,a4paper]{IEEEtran}
\usepackage{times,amsmath,epsfig}
\title{Low-Complexity Adaptive Set-Membership Reduced-rank LCMV Beamforming }
\author{%
{Lei Wang, Rodrigo C. de Lamare}%
\vspace{1.6mm}\\
\fontsize{10}{10}\selectfont\itshape
Communications Research Group, Department of Electronics\\
University of York, York YO10 5DD, UK\\
\fontsize{9}{9}\selectfont\ttfamily\upshape
lw517@ohm.york.ac.uk\\
rcdl500@ohm.york.ac.uk}
\begin{document}
\maketitle
\begin{abstract}
This paper proposes a new adaptive algorithm for the implementation
of the linearly constrained minimum variance (LCMV) beamformer. The
proposed algorithm utilizes the set-membership filtering (SMF)
framework and the reduced-rank joint iterative optimization (JIO)
scheme. We develop a stochastic gradient (SG) based algorithm for
the beamformer design. An effective time-varying bound is employed
in the proposed method to adjust the step sizes, avoid the
misadjustment and the risk of overbounding or underbounding.
Simulations are performed to show the improved performance of the
proposed algorithm in comparison with existing full-rank and
reduced-rank methods.
\end{abstract}

%
\section{Introduction}
Adaptive beamforming technology has received considerable attention
for several decades and found widespread applications in radar,
sonar and wireless communications \cite{Jian}. The optimal linearly
constrained minimum variance (LCMV) beamformer \cite{Frost} which
minimizes the array output power while maintaining the array
response on the direction of the desired signal, is the most
well-known beamformer. However, the computation of the inverse of
the received data covariance matrix and the requirement of the
knowledge of the cross-correlation vector present difficulties for
its implementation.

A number of adaptive algorithms have been reported for the
implementation of the LCMV beamformer \cite{Jian}, \cite{Trees}.
Among them the stochastic gradient (SG) algorithm \cite{Haykin},
\cite{Lamare3} is popular due to its simplicity and low complexity.
The drawback of the SG algorithm is that the convergence depends on
the eigenvalue spread of the received data covariance matrix. This
condition could be worse when the number of elements in the filter
is large since it requires a large amount of snapshots to reach the
steady-state.

Reduced-rank signal processing was introduced to provide a way out
of this dilemma \cite{Haimovich}-\cite{Lamare}. The reduced-rank
schemes project the received vector onto a lower dimensional
subspace and perform the filter optimization within this subspace.
The advantages are their fast convergence properties and enhanced
tracking performance when compared with full-rank schemes operating
with a large number of parameters in the filter for the beamformer
design. The reduced-rank algorithms range from the auxiliary vector
filtering (AVF) \cite{Pados}, the multistage Wiener filter (MSWF)
\cite{Honig}, to the joint iterative optimization (JIO) based works
reported in \cite{Lamare4}, \cite{Lamare}. Despite the improved
convergence and tracking performance achieved with the existing
reduced-rank methods, they have to afford the heavy computational
load as a tradeoff.

In this paper, we introduce a new LCMV reduced-rank algorithm based
on the JIO scheme. The proposed algorithm employs the set-membership
filtering (SMF) technique \cite{Diniz}, \cite{Huang} to reduce the
computational complexity significantly without the convergence speed
reduction and the performance loss. The SMF specifies a bound on the
magnitude of the estimation error (or the array output) and uses the
data-selective updates to adjust parameters according to this
predetermined bound. It involves two steps: $1)$ information
evaluation and $2$) parameters adaptation. If the parameter update
does not occur frequently, and the information evaluation does not
require much complexity, the overall computational cost can be saved
substantially. The current SMF algorithms (see \cite{Lamare2} and
the reference therein) pay attention to the full-rank parameter
estimation. Considering the fact that the reduced-rank algorithms
exhibit superior performance over the full-rank methods
\cite{Honig}, \cite{Lamare}, \cite{Chen}, it motivates the
deployment of the SMF mechanism to the reduced-rank scheme to
guarantee the good performance with low complexity. We employ the
SMF technique with the reduced-rank JIO scheme for the LCMV
beamformer design and develop the SG based algorithm for
implementation. Compared with the work reported in \cite{Diniz},
\cite{Lamare2}, the devised scheme consists of a bank of full-rank
adaptive filters, which constitutes the projection matrix, and an
adaptive reduced-rank filter that operates at the output of the bank
of full-rank filters. It provides an iterative exchange of
information between the projection matrix and the reduced-rank
filter and thus leads to improved convergence and tracking
performance. Compared with the JIO based method (with the fixed step
sizes) in \cite{Lamare}, the proposed algorithm uses the SMF
technique to adjust the step sizes for the updates of the projection
matrix and the reduced-rank weight vector, and thus has an
attractive tradeoff between the convergence rate and misadjustment.
Furthermore, a time varying bound is incorporated in the proposed
algorithm to avoid the risk of overbounding and underbounding in
dynamic scenarios \cite{Lamare2}, and to improve the performance
with a small number of update.

The remaining of this paper is organized as follows: we outline a
system model for beamforming and present the problem statement in
Section 2. Section 3 derives the proposed adaptive reduced-rank
algorithm. Simulation results are provided and discussed in Section
4, and conclusions are drawn in Section 5.

\section{System Model and Problem Statement}

\subsection{System Model}

Let us suppose that $q$ narrowband signals impinge on a uniform
linear array (ULA) of $m$ ($q \leq m$) sensor elements. The sources
are assumed to be in the far field with directions of arrival (DOAs)
$\theta_{0}$,\ldots,$\theta_{q-1}$. The received vector $\boldsymbol
x(i)\in\mathcal C^{m\times 1}$ at the $i$th snapshot can be modeled
as
\begin{equation} \label{1}
\centering {\boldsymbol x}(i)={\boldsymbol A}({\boldsymbol
{\theta}}){\boldsymbol s}(i)+{\boldsymbol n}(i),~~~ i=1,\ldots,N
\end{equation}
where
$\boldsymbol{\theta}=[\theta_{0},\ldots,\theta_{q-1}]^{T}\in\mathcal{R}^{q
\times 1}$ is the DOAs, ${\boldsymbol A}({\boldsymbol
{\theta}})=[{\boldsymbol a}(\theta_{0}),\ldots,{\boldsymbol
a}(\theta_{q-1})]\in\mathcal{C}^{m \times q}$ composes the steering
vectors ${\boldsymbol a}(\theta_{k})=[1,e^{-2\pi
j\frac{d}{\lambda_{c}}cos{\theta_{k}}},\ldots,e^{-2\pi
j(m-1)\frac{d}{\lambda_{c}}cos{\theta_{k}}}]^{T}\in\mathcal{C}^{m
\times 1},~~~(k=0,\ldots,q-1)$, where $\lambda_{c}$ is the
wavelength and $d=\lambda_{c}/2$ is the inter-element distance of
the ULA, and to avoid mathematical ambiguities, the steering vectors
$\boldsymbol a(\theta_{k})$ are considered to be linearly
independent. ${\boldsymbol s}(i)\in \mathcal{C}^{q\times 1}$ is the
source data, ${\boldsymbol n}(i)\in\mathcal{C}^{m\times 1}$ is the
white Gaussian noise, $N$ is the observation size of snapshots, and
$(\cdot)^{T}$\ stands for the transpose. The output of a narrowband
beamformer is
\begin{equation} \label{2}
\centering y(i)={\boldsymbol w}^H{\boldsymbol x}(i),
\end{equation}
where ${\boldsymbol
w}=[w_{1},\ldots,w_{m}]^{T}\in\mathcal{C}^{m\times 1}$ is the
complex weight vector, and $(\cdot)^{H}$ stands for the Hermitian
transpose.

\subsection{Problem Statement}

The optimal LCMV filter for beamforming can be computed by solving
the following optimization problem
\begin{equation}\label{3}
\begin{split}
&\textrm{minimize}~~~E[|\boldsymbol w^{H}\boldsymbol x(i)|^{2}]\\
&\textrm{subject~to}~~~{\boldsymbol w}^{H}{\boldsymbol
a}(\theta_{0})=\gamma,
\end{split}
\end{equation}
where $\boldsymbol a(\theta_0)$ denotes the steering vector of the
desired signal and $\gamma$ is a constant with respect to the
constraint. The weight solution is
\begin{equation}\label{4}
\boldsymbol w=\frac{\gamma\boldsymbol R^{-1}\boldsymbol
a(\theta_{0})}{\boldsymbol a^{H}(\theta_{0})\boldsymbol
R^{-1}\boldsymbol a(\theta_{0})},
\end{equation}
where $\boldsymbol R=E[\boldsymbol x(i)\boldsymbol
x^{H}(i)]\in\mathcal C^{m\times m}$ is the received vector
covariance matrix. The complexity of the weight computation is high
due to the existence of the covariance matrix inverse in (\ref{4}).
The SG algorithm \cite{Haykin} can be used to estimate $\boldsymbol
w(i)$ with low complexity but suffers from the slow convergence,
especially when the array size is large.

The most important feature of the reduced-rank algorithms is to
perform the dimensionality reduction and retain the key information
of the original signal in the reduced-rank received vector, which is
\begin{equation}\label{5}
\bar{\boldsymbol x}(i)=\boldsymbol T_r^H\boldsymbol x(i),
\end{equation}
where $\boldsymbol T_r\in\mathcal C^{m\times r}$ denotes the
projection matrix that is structured as a bank of $r$ full-rank
filters $\boldsymbol t_j=[t_{1,j},\ldots, t_{m,j}]^T\in\mathcal
C^{m\times1}$, ($j=1,\ldots,r$), $r$ is the rank number, and
$\bar{\boldsymbol x}(i)\in\mathcal C^{r\times1}$ is the reduced-rank
received vector. In what follows, all $r$-dimensional quantities are
denoted by an over bar. An adaptive reduced-rank filter
$\bar{\boldsymbol w}=[\bar{w}_1,\ldots,\bar{w}_r]^T\in\mathcal
C^{r\times1}$ is followed to produce the output
\begin{equation}\label{6}
y(i)=\bar{\boldsymbol w}^H\bar{\boldsymbol x}(i).
\end{equation}

The main concern left to us is how to effectively design and
calculate the projection matrix. The popular reduced-rank schemes
include AVF \cite{Pados}, MSWF \cite{Honig}, and JIO \cite{Lamare2}.
The SG type algorithm can be employed in these schemes to estimate
$\bar{\boldsymbol w}$. However, it is difficult to set the step size
for the existing methods to achieve a satisfactory tradeoff between
the fast convergence and the misadjustment in dynamic scenarios.
Besides, the generation of the projection matrix is still a
complicated task that increases the computational cost.

\section{Proposed Reduced-rank SMF Algorithm}

In this section, we introduce a new adaptive reduced-rank algorithm
based on the JIO scheme and employ the SMF technique with the
time-varying bound to realize the data-selective updates for the
beamformer design. It should be remarked that the JIO scheme is
selected here due to its improved performance and relatively simple
implementation over the AVF and the MSWF ones \cite{Lamare}.

\subsection{Proposed Set-Membership Scheme}

In the existing full-rank SMF scheme \cite{Huang}-\cite{Diniz2}, the
filter $\boldsymbol w$ is designed to achieve a predetermined or
time-varying bound on the magnitude of the estimation error (or the
array output). This bound can be regarded as a constraint on the
filter design, which performs the updates for certain received data,
namely, the data-selective updates. For the reduced-rank JIO scheme
with the SMF technique, we need to take both the projection matrix
$\boldsymbol T_r$ and the reduced-rank weight vector
$\bar{\boldsymbol w}$ into consideration due to the feature of their
joint iterative exchange of information. Let $\Theta(i)$ represent
the set containing all the pairs of $\{\boldsymbol T_r(i),
\bar{\boldsymbol w}(i)\}$ for which the corresponding array output
at time instant $i$ is upper bounded in magnitude by a time-varying
bound $\delta(i)$, yields,
\begin{small}
\begin{equation}\label{7}
\Theta(i)=\bigcap_{\big(s_0(i),\boldsymbol x(i)\big)\in\mathcal
{\boldsymbol S}}\Big\{\bar{\boldsymbol w}\in\mathcal C^{r\times
1},\boldsymbol T_r\in\mathcal C^{m\times
r}:|y(i)|^2\leq\delta^2(i)\Big\},
\end{equation}
\end{small}
where $s_0(i)$ denotes the transmitted data of the desired user from
$\theta_0$, $\mathcal {\boldsymbol S}$ is the set of all possible
data pairs ($s_0(i),\boldsymbol x(i)$), and the set $\Theta(i)$ is
referred to as the \textit{feasibility} set. The pairs of
$\{\boldsymbol T_r(i), \bar{\boldsymbol w}(i)\}$ in the set satisfy
the constraint $|y(i)|^2\leq\delta^2(i)$. In practice, $\mathcal
{\boldsymbol S}$ cannot be traversed allover. A larger space of the
data pairs provided by the observations leads to a smaller
feasibility set. That is, as the number of data pairs increases,
there are fewer pairs of $\{\boldsymbol T_r(i), \bar{\boldsymbol
w}(i)\}$ that can be found to satisfy the constraint. Under this
condition, we define the exact \textit{membership} sets $\Psi(i)$ to
be the intersection of the \textit{constraint} sets over the time
instants $i=1,\ldots, N$, i.e.,
\begin{equation}\label{8}
\Psi(i)=\bigcap_{l=1}^i\mathcal {H}_l,
\end{equation}
where $\mathcal H_l=\{\bar{\boldsymbol w}\in\mathcal
C^{r\times1},\boldsymbol T_r\in\mathcal C^{m\times
r}:|y(i)|^2\leq\delta^2(i)\}$ is the constraint set to provide
information for the membership set to construct a set of solutions.
It is clear that the feasibility set $\Theta(i)$ is a limiting set
of the exact membership set $\Psi(i)$ in practice. The two sets will
be equal if the data pairs traverse $\mathcal {\boldsymbol S}$
completely.

The proposed SMF scheme introduces the principle of set-membership
into reduced-rank signal processing and inherits the advantage of
the joint optimization between the projection matrix and the
reduced-rank filter. It updates the parameter vectors such that they
will always belong to the feasibility set. Note that the
time-varying bound $\delta(i)$ has to be chosen appropriately in
order to devise an effective algorithm and avoid the risk of
overbounding or underbounding. The selection of the bound will be
given in the following part.

\subsection{Proposed Reduced-Rank SMF Algorithm}
In this section, we use the data-selective updates of the proposed
SMF scheme in the reduced-rank JIO based algorithm. For the LCMV
beamformer design, the proposed algorithm can be derived according
to the minimization of the following cost function
\begin{equation}\label{9}
\begin{split}
&\textrm{minimize}~~E\big[|\bar{\boldsymbol
w}^H\boldsymbol T_r^H{\boldsymbol x}(i)|^2\big]\\
&\textrm{subject~to}~~\bar{\boldsymbol w}^H\bar{\boldsymbol
a}(\theta_0)=\gamma,~~\textrm{and}~~|\bar{\boldsymbol
w}^H\boldsymbol T_r^H{\boldsymbol x}(i)|^2=g^2(i),
\end{split}
\end{equation}
where $\bar{\boldsymbol x}(i)=\boldsymbol T_r^H\boldsymbol x(i)$ is
the reduced-rank received vector defined in (\ref{5}),
$\bar{\boldsymbol a}(\theta_0)=\boldsymbol T_r^H\boldsymbol
a(\theta_0)$ is the reduced-rank steering vector of the desired
user, and $g(i)$ is a coefficient that determines a set of estimates
within the constraint set $\mathcal H_i$ and $|g(i)|\leq\delta(i)$.
The cost function in (\ref{9}) depends on the projection matrix
$\boldsymbol T_r$ and the reduced-rank filter $\bar{\boldsymbol w}$.
The solution of (\ref{9}) should be encompassed in the feasibility
set in (\ref{7}) and the pairs of $\{\boldsymbol T_r(i),
\bar{\boldsymbol w}(i)\}$ should satisfy the constraint set
$\mathcal H_i=\{\bar{\boldsymbol w}, \boldsymbol
T_r:|y(i)|^2\leq\delta^2(i)\}$.

In order to solve the optimization problem in (\ref{9}), using the
method of Lagrange multipliers \cite{Haykin} and considering the
constraint $\bar{\boldsymbol w}^H\bar{\boldsymbol
a}(\theta_0)=\gamma$, we have
\begin{equation}\label{10}
J(\bar{\boldsymbol w},\boldsymbol T_r)= E\big[\bar{\boldsymbol
w}^H\boldsymbol T_r^H{\boldsymbol x}(i){\boldsymbol
x}^H(i)\boldsymbol T_r\bar{\boldsymbol w}\big]+2\lambda
Re\big[\bar{\boldsymbol w}^H\bar{\boldsymbol
a}(\theta_0)-\gamma\big]
\end{equation}
where $\lambda$ is the Lagrange multiplier and $Re[\cdot]$ selects
the real part of the quantity. Note that the constraint
$|\bar{\boldsymbol w}^H\boldsymbol T_r^H{\boldsymbol
x}(i)|^2=g^2(i)$ is not included in (\ref{10}). This is because a
point estimate can be obtained from (\ref{10}) whereas the bounded
constraint  determines a set of $\{\boldsymbol T_r, \bar{\boldsymbol
w}\}$. We use the constraint with respect to $\boldsymbol
a(\theta_0)$ to determine one solution and to expand it to a
hyperplane by using the bounded constraint.

We use the SG type algorithm to update the parameter vectors.
Specifically, assuming $\bar{\boldsymbol w}$ is known, computing the
\textit{instantaneous} gradient of (\ref{10}) with respect to
$\boldsymbol T_r$, equating it to a zero matrix and solving for
$\lambda$, we have
\begin{equation}\label{11}
\boldsymbol T_r(i+1)=\boldsymbol
T_r(i)-\mu_{T}y^{\ast}(i)\big[\boldsymbol I-\boldsymbol
a(\theta_0)\boldsymbol a^H(\theta_0)\big]\boldsymbol
x(i)\bar{\boldsymbol w}^H(i),
\end{equation}
where $\mu_{T}$ is the step size for the update of the projection
matrix and $\boldsymbol I$ is the corresponding identity matrix.

Assuming $\boldsymbol T_r$ is known, taking the instantaneous
gradient of (\ref{10}) with respect to $\bar{\boldsymbol w}$,
equating it to a null vector and solving for $\lambda$, we get
\begin{equation}\label{12}
\bar{\boldsymbol w}(i+1)=\bar{\boldsymbol
w}(i)-\mu_{\bar{w}}y^{\ast}(i)\big[\boldsymbol
I-\frac{\bar{\boldsymbol a}(\theta_0)\bar{\boldsymbol
a}^H(\theta_0)}{\bar{\boldsymbol a}^H(\theta_0)\bar{\boldsymbol
a}(\theta_0)}\big]\bar{\boldsymbol x}(i)
\end{equation}
where $\mu_{\bar{w}}$ is the step size for the update of the
reduced-rank weight vector.

The updates of the projection matrix and the reduced-rank weight
vector may spend a long time or terminate suddenly due to the
misadjustment if the step sizes cannot be set suitably. The second
constraint in (\ref{9}) can be used to set a bound on the power of
the array output for adjusting the step size to avoid these
problems. The predetermined bound \cite{Diniz}, \cite{Huang} in the
SMF technique makes contributions to this topic. However, we cannot
often determine the bound accurately since there is usually
insufficient knowledge about the underlying system, especially when
the scenario is changing. In such cases, a predetermined bound
always has the risk of underbounding or overbounding, both of which
result in the performance degradation. It motivates us to introduce
a time-varying bound to circumvent the aforementioned problems.

In the proposed algorithm, we use a time-varying bound in the second
constraint of (\ref{9}) to offer a good tradeoff between the
convergence rate and the misadjustment by adjusting the step size
values automatically following the time instant. Under this
condition, the update is performed only if the constraint
$|\bar{\boldsymbol w}^H\bar{\boldsymbol x}(i)|^2=g^2(i)$ cannot be
satisfied. This scheme provides data-selective updates (updates with
respect to certain snapshots) for the filter design and thus reduces
the computational complexity compared with the existing reduced-rank
algorithms (updates for all the snapshots). Specifically, by
substituting the update equations of (\ref{11}) and (\ref{12}) into
the constraint on the time-varying bound, respectively, we obtain
{\small
\begin{equation}\label{13}
\mu_{T}(i)=\left\{ \begin{array}{ccc}
                  \frac{1-\frac{\delta(i)}{|y(i)|}}{\bar{\boldsymbol w}^H(i)\bar{\boldsymbol w}(i)\boldsymbol x^H(i)[\boldsymbol I-\boldsymbol a(\theta_0)\boldsymbol a^H(\theta_0)]{\boldsymbol x}(i)}&\textrm {if}|y(i)|^2\geq\delta^2(i)\\
                    &\\
                  0 & \textrm {otherwise}\\
                  \end{array}\right.
\end{equation}
and
\begin{equation}\label{14} \mu_{\bar{w}}(i)=\left\{
\begin{array}{ccc}
                 \frac{1-\frac{\delta(i)}{|y(i)|}}{\bar{\boldsymbol x}^H(i)[\boldsymbol I-\frac{\bar{\boldsymbol a}(\theta_0)\bar{\boldsymbol a}^H(\theta_0)}{\bar{\boldsymbol a}^H(\theta_0)\bar{\boldsymbol a}(\theta_0)}]\bar{\boldsymbol
                 x}(i)} & \textrm {if}~|y(i)|^2\geq\delta^2(i)\\
                 & \\
                 0 & \textrm{otherwise}.\\
                 \end{array}\right.
\end{equation}}
The only challenge left to us now is how to select the
time-varying bound $\delta(i)$ in the constraint to make the
proposed algorithm work effectively. We introduce a
parameter-dependent bound (PDB) that was reported in \cite{Guo} to
update the reduced-rank weight vector for capturing the desired user
and suppressing the interference and noise., that is
\begin{equation}\label{15}
\delta(i)=\beta\delta(i-1)+(1-\beta)\sqrt{\alpha\|\boldsymbol
T_r(i)\bar{\boldsymbol w}(i)\|^2\hat{\sigma}_n^2(i)},
\end{equation}
where $\beta$ is a forgetting factor that should be set to guarantee
an appropriate time-averaged estimate of the evolutions of the
weight vector $\boldsymbol w(i)$, which is given by $\boldsymbol
w(i)=\boldsymbol T_r(i)\bar{\boldsymbol w}(i)$, $\|\boldsymbol
T_r(i)\bar{\boldsymbol w}(i)\|^2\hat{\sigma}_n^2(i)$ is the variance
of the inner product of the weight vector with $\boldsymbol n(i)$
that provides information on the evolution of $\boldsymbol w(i)$,
$\alpha$ is a tuning coefficient with $\alpha>1$ \cite{Guo}, and
$\hat{\sigma}_n^2(i)$ is an estimate of the noise power. This
time-varying bound provides a smoother evolution of the weight
vector trajectory and thus avoids too high or low values of the
squared norm of $\boldsymbol w(i)$.

Until now, we finish the derivation of the proposed algorithm, which
is called JIO-SM-SG. A summary of the proposed method is given in
Table \ref{tab:JIO-SM-SG}, where the initialization procedure is
important to start the update. From (\ref{11}) and (\ref{12}), the
projection matrix and the reduced-rank weight vector depend on each
other, which provides an iterative exchange of information between
each other and thus leads to an improved performance. The SMF
technique is employed in the proposed algorithm and the time-varying
bound is calculated with the incoming of the received vector. It
encompasses the pairs of the filter design $\{\boldsymbol T_r(i),
\bar{\boldsymbol w}(i)\}$ in the feasibility set $\Theta(i)$ that
satisfies the constraint $|y(i)|^2\leq\delta^2(i)$. The filter
updates are not performed for each time instant $i$ but only when
this constraint cannot reach (i.e., data-selective updates). The
computational complexity is reduced significantly and the
convergence rate is further increased compared with the existing
reduced-rank methods. The proposed algorithm combines the positive
features of the reduced-rank JIO scheme and time-varying based SMF
technique for the implementation of the LCMV beamformer design.
\begin{table}[!t]
\centering
    \caption{THE PROPOSED JIO-SM-SG ALGORITHM}     
    \label{tab:JIO-SM-SG}
    \begin{small}
        \begin{tabular}{|l|}
\hline
\bfseries {Initialization:}\\
\hline
${\boldsymbol T}_r(1)=[{\boldsymbol I}_{r\times r}~\boldsymbol 0_{r\times (m-r)}]^T$;\\
${\bar{\boldsymbol w}}(1)=\boldsymbol T_r^H(1)\boldsymbol a(\theta_0)/(\|\boldsymbol T_r^H(1)\boldsymbol a(\theta_0)\|^2)$;\\
$\mu_{T}(1)$~\textrm{and}~$\mu_{\bar{w}}(1)=\textrm{small positive
values}$.\\
\hline
\bfseries {For each time instant} $i=1,\ldots, N$\\
\hline $\bar{\boldsymbol x}(i)=\boldsymbol T_r^H(i)\boldsymbol
x(i)$;~~~$y(i)=\bar{\boldsymbol w}^H(i)\bar{\boldsymbol x}(i)$;~~~
$\delta(i)$~\textrm{in}~~~~~~(\ref{15})\\
~~~~~~\bfseries {if} $|y(i)|^2\geq\delta^2(i)$\\
~~~~~~~~~~~~~$\mu_{T}(i)$~\textrm{in}~~~~~~~~~~~~~~~~~~~~~~~~~~~~~~~~~~~~~~~~~~~~(\ref{13})\\
~~~~~~~~~~~~~$\boldsymbol T_r(i+1)$~\textrm{in}~~~~~~~~~~~~~~~~~~~~~~~~~~~~~~~~~~~~~~~(\ref{11})\\
~~~~~~~~~~~~~$\bar{\boldsymbol a}(\theta_0)=\boldsymbol
T_r^H(i)\boldsymbol a(\theta_0)$\\
~~~~~~~~~~~~~$\mu_{\bar{w}}(i)$~\textrm{in}~~~~~~~~~~~~~~~~~~~~~~~~~~~~~~~~~~~~~~~~~~~~(\ref{14})\\
~~~~~~~~~~~~~$\bar{\boldsymbol w}(i+1)$~\textrm{in}~~~~~~~~~~~~~~~~~~~~~~~~~~~~~~~~~~~~~~~~(\ref{12})\\
~~~~~~\bfseries {else}\\
~~~~~~~~~~~~~$\boldsymbol T_r(i+1)=\boldsymbol T_r(i)$;~~~$\bar{\boldsymbol w}(i+1)=\bar{\boldsymbol w}(i)$\\
~~~~~~\bfseries {end}\\
\hline
    \end{tabular}
    \end{small}
\end{table}

\section{Simulation Results}

We evaluate the performance of the proposed JIO-SM-SG algorithm and
compared it with those of the existing methods, i.e., full-rank SG
\cite{Haykin}, full-rank SG with SMF \cite{Lamare2}, AVF
\cite{Pados}, MSWF-SG \cite{Honig}, and reduced-rank JIO-SG
\cite{Lamare}. We assume that the DOA of the desired user is known
by the receiver. In each experiment, a total of $K=1000$ runs are
carried out to obtain the curves. We use the BPSK and set the input
SNR$=10$ dB and INR$=30$ dB. Simulations are performed by an ULA
containing $m=64$ sensor elements with half-wavelength interelement
spacing.

\begin{figure}[h]
    \centerline{\psfig{figure=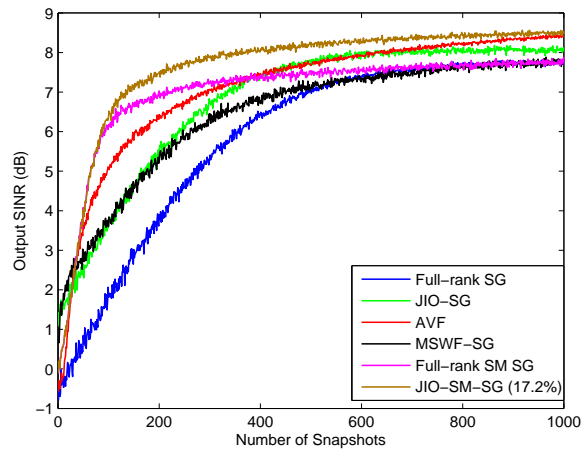,height=64.54mm} }
    \caption{Output SINR versus the
number of snapshots}
    \label{fig:iswcs102_final}
\end{figure}

In Fig. \ref{fig:iswcs102_final}, there are $q=25$ users, including
one desired user in the system. The step size values are initialized
by $\mu_{T}(1)=\mu_{\bar{w}}(1)=0.05$. We set $\alpha=22$,
$\beta=0.99$, and the rank $r=5$. This experiment exhibits that the
output SINR values of the existing and the proposed algorithms
increase to the steady-state as the increase of the snapshots (time
index). The proposed algorithm shows faster convergence and better
performance over the existing methods. The step size values are
adapted to ensure the fast convergence rate without the risk of
misadjustment for the proposed algorithm. Due to the data-selective
updates, the proposed algorithm could reduce the computational load
significantly as it only requires $17.2\%$ updates ($172$ updates
for $1000$ snapshots), which is significantly lower than those of
the existing reduced-rank methods (normally $1000$ updates for
$1000$ snapshots).

\begin{figure}[h]
    \centerline{\psfig{figure=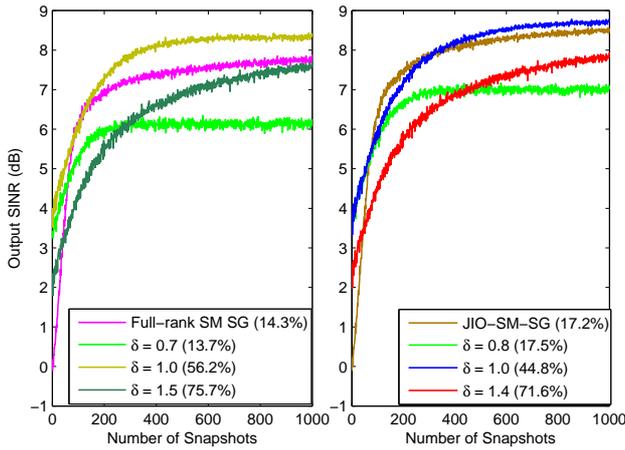,height=64.54mm} }
    \caption{Output SINR versus the
number of snapshots for (a) full-rank SM SG algorithm; (b) JIO-SM-SG
algorithm}
    \label{fig:iswcs101_final}
\end{figure}

Fig. \ref{fig:iswcs101_final}, which includes two experiments, shows
the effectiveness of the proposed algorithm with the time-varying
bound. The scenario is the same as that in Fig.
\ref{fig:iswcs102_final}. We compare the full-rank (Fig.
\ref{fig:iswcs101_final} (a)) and proposed (Fig.
\ref{fig:iswcs101_final} (b)) algorithms with the fixed and
time-varying bounds, respectively. In Fig. \ref{fig:iswcs101_final}
(a), the one with the fixed bound $\delta=1.0$ shows better
performance than that with the time-varying bound but increases the
computational cost as a tradeoff ($56.2\%$). The algorithms with
higher ($\delta=1.5$) or lower ($\delta=0.7$) bounds exhibit worse
convergence and steady-state performance. The same result can be
found in Fig. \ref{fig:iswcs101_final} (b) for the proposed
algorithm. It indicates that the time-varying bound is capable of
improving the performance of the proposed JIO-SM-SG algorithm, while
realizing the data-selective updates to reduce the computation
complexity.

\section{Conclusion}

In this paper, we employed the time-varying bound SMF technique in
the reduced-rank JIO scheme and developed a new adaptive algorithm
for the LCMV beamformer. The proposed algorithm retained the
positive feature of the iterative exchange of information between
the projection matrix and the reduced-rank weight vector, and used
data-selective updates to adjust parameters that satisfy the
constraints of the LCMV cost function. The variable step sizes in
the proposed algorithm provided a way to circumvent the problem
between the convergence and misadjustment. It achieved superior
performance, especially in large array scenarios, with relatively
low complexity.

\end{document}